\documentclass[twoside,11pt,letterpaper]{article}

\usepackage{titlesec}
\usepackage{latexsym}
\usepackage{theorem}
\usepackage{fancyhdr}
\usepackage{amssymb}
\usepackage{amsmath}

\renewenvironment{abstract}[1]{\begin{minipage}[t]{\textwidth}
\mbox{\Large #1} \hspace{-0.5em}}{\end{minipage}}

\newcommand{\nn}{\nonumber}

\newcommand{\beq}{\begin{equation}}
\newcommand{\eeq}{\end{equation}}
\newcommand{\beqarr}{\begin{eqnarray}\vspace{1em}}
\newcommand{\eeqarr}{\vspace{1em}\end{eqnarray}}

\newcommand{\bdefinition}{\begin{definition}}
\newcommand{\edefinition}{\end{definition}} 
\newcommand{\btheo}{\begin{theorem}\vspace{1em}}
\newcommand{\etheo}{\vspace{1em}\end{theorem}}

\newcommand{\laction}{\triangleright}

\newcommand{\tensor}{\otimes}

\newcommand{\id}{\textrm{id}}
\newcommand{\coproduct}{\Delta}
\newcommand{\counit}{\epsilon}
\newcommand{\one}{\mathbf{1}}
\newcommand{\antipode}{\textrm{S}}

\newcommand{\algebra}{\mathfrak{A}}
\newcommand{\hopfalgebra}{\mathcal{H}}

\newcommand{\liealgebra}{\mathfrak{g}}

\newcommand{\field}{\mathbf{K}}
\newcommand{\X}{\mathfrak{X}}
\newcommand{\V}{\mathfrak{V}}

\newcommand{\R}{\mathbb{R}}
\newcommand{\quasiR}{\mathcal{R}}
\newcommand{\twist}{\mathcal{F}}
\newcommand{\leftcrossproduct}{>\!\!\!\triangleleft}
\newcommand{\unit}{\eta}

\newcommand{\product}{\mu}

\theoremstyle{change}

\theoremheaderfont{\scshape}

\newtheorem{definition}{Definition}[section]
\newtheorem{theorem}[definition]{Theorem}

\newtheorem{proposition}[definition]{Proposition}
\newtheorem{defproposition}[definition]{Definition-Proposition}

\titlespacing{\section}{0em}{2em}{2em}
\titlespacing{\subsection}{0em}{2em}{2em}

\titleformat{\section}
  {\sc \large}
  {\thesection}
  {1em}
  {}
\titleformat{\subsection}
  {\sc}
  {\thesubsection}
  {1em}
  {}

\numberwithin{equation}{section}

\begin{document}


\title{
\begin{flushright}
	\small \scshape LMU-ASC 45/06 \\
	\small \scshape MPP-2006-84\\[1.7em]
\end{flushright}
\scshape Vector Field Twisting of Lie-Algebras}
\author{\renewcommand{\thefootnote}{\arabic{footnote}} \scshape Florian Koch\footnotemark[1] \\ 
\rule{20mm}{0.2mm} \\[0.5em]
\small \scshape Arnold Sommerfeld Center for Theoretical Physics\\[-0.4em]
\small \scshape Ludwig-Maximilians-Universit\"at M\"unchen\\[-0.4em]
\small \scshape Theresienstra{\ss}e 37, 80333 M\"unchen, Germany\\[0.8em]
\small \scshape Werner Heisenberg Institut\\[-0.4em]
\small \scshape Max-Planck-Institut f\"ur Physik\\[-0.4em]
\small \scshape F\"ohringer Ring 6, 80805 M\"unchen, Germany\\[0.5em]
\rule{20mm}{0.2mm} \\[0.5em]}
\date{}

\footnotetext[1]{koch@theorie.physik.uni-muenchen.de}

\maketitle


\begin{abstract}{I}n quantum groups coproducts of Lie-algebras are twisted in terms of 
generators of the corresponding universal enveloping algebra. If representations are considered, 
twists also serve as starproducts that accordingly quantize representation spaces. 
In physics, requirements turn out to be the other way around. Physics comes up with 
noncommutative spaces in terms of starproducts that miss a suiting quantum symmetry. 
In general the classical limit is known, i.e. there exists a representation of 
the Lie-algebra on a corresponding finitely generated commutative space. In this setup
quantization can be considered independently from any representation theoretic 
issue. We construct an algebra of vector fields from a left cross-product algebra of the 
representation space and its Hopf-algebra of momenta. The latter can always be defined. 
The suitingly devided cross-product algebra is then lifted to a Hopf-algebra that 
carries the required genuine structure to accomodate a matrix representation
of the universal enveloping algebra as a subalgebra. We twist the Hopf-algebra of vector 
fields and thereby obtain the desired twisting of the Lie-algebra. 
Since we twist with vector fields and not with generators of the Lie-algebra, this is 
the most general twisting that can possibly be obtained. In other words, we push starproducts 
to twists of the desired symmetry algebra and to this purpose solve the problem of 
turning vector fields into a Hopf-algebra. We give some genuine example.
\end{abstract}

\setcounter{footnote}{1}
\thispagestyle{empty}



\newpage

\section{Introduction}

Studies of quantum groups require for a considerable mathematical framework that historically
caused the topic to be turned into a mathematical field on its own. As a consequence it then
naturally followed its own mathematical interests - apart from actual physical requirements. 
In quantum groups deformations of a Lie-algebra $\liealgebra$ are considered in terms of its 
universal enveloping algebra $U(\liealgebra)$. Coproducts of $U(\liealgebra)$ are deformed by 
conjugation with quasitriangular structures  $\quasiR \in U(\liealgebra) \tensor U(\liealgebra)$ 
or twists $\twist \in U(\liealgebra) \tensor U(\liealgebra)$. The noncocommutative coproduct 
of the deformed version of the universal enveloping algebra $U(\liealgebra)$ dually implies a 
noncommutative structure on representation space. As an example see \cite{Blohmann:2002wx}. 
Thus within the standard workflow of quantum groups, symmetry algebras are first deformed and 
represented afterwards. Physics, however, requires for the opposit procedure. Theories and 
models come down with noncommutative spaces, as canonical spacetime in \cite{Chu:1998qz,Schomerus:1999ug,Seiberg:1999vs}, that miss the corresponding quantum symmetry. 
In most cases the classical limit exists, i.e. there exists a representation of $\liealgebra$ 
on a finitely generated commutative space. The task at hand is to find the corresponding 
deformation of the symmetry algebra. But quantum groups do not provide the required techniques. 
It thus takes quite a time until such quantizations are found - if they are found at all. 
For the case of canonical commutation relations these were constructed in \cite{Oeckl:2000eg,Chaichian:2004za,Wess:2003da,Koch:2004ud}. While twists can be used as 
starproducts, the opposit only holds for some specific exceptions. This is the standard situation 
in physics. Quite often it has been observed that quantization requires for some enhancement 
of the symmetry algebra \cite{Zachos:2001ux}. For example, the well-known $\kappa$-deformation 
of the Poincar\'e algebra \cite{Lukierski:1992dt,Lukierski:1991ff,Lukierski:1991pn} cannot be reduced 
to that of the Lorentz-algebra alone. The algebra of momenta is a vital component of this 
deformation. The mathematical setup to this example had been provided by \cite{Majid:1994cy}. 
The same holds for the mentioned $\theta$-deformation of the Poincar\'e algebra for canonical 
commutation relations. Obviously only those very specific deformations can merely be performed 
\emph{within} the symmetry algebra, that are ruled by a quasitriangular structure $\quasiR$. But these 
only provide quantum spaces with quadratic commutation relations. We can thus observe the 
\emph{physical reason} why $\kappa$- and $\theta$-deformations required for some algebraic 
enhancement: The deformation parameter carries a physical dimension. Thus while the mathematical 
workflow restricted to a single version of quantum spaces, that turned out quite unhandy for 
physical applications, physics itself came up with deformations beyond this setup. 
And mathematics, as often, delivered an explanation afterwards. The universal enveloping 
algebra of a Lie-algebra is obviously not large enough in order to 
perform most general quantizations of its coproducts. The authors of \cite{Lukierski:2005fc,
Lukierski:2005jb}, \cite{Borowiec:2005bq} incorporated this idea and used the Poincar\'e algebra as 
a whole in order to obtain more general twistings. They receive quantum spaces with quadratic as 
well as Lie-algebra valued commutation relations. Here we want to push this a little further.
Within another example of physics, phase space deformations were considered in order to obtain
high energy motivated minimal uncertainty models \cite{Kempf:1998em,Kempf:1998gk,Kempf:1997sr,Kempf:1994qp}.
The author speculates that the deformation of a corresponding Poincar\'e-algebra might be obtained by 
the use of the phase space algebra itself. In contrast to this, the authors of \cite{Jambor:2004kc} 
formulate starproducts in terms of vector fields. Vector fields are most fundamental objects of 
differential geometry and Lie-algebras themselves describe nothing else than the currents on curved 
manifolds. Apart from this, there is a close relation between noncommutative geometry and quantization 
over curved spaces. In this respect vector fields also played a crucial role for noncommutative
gravity \cite{Aschieri:2005yw,Aschieri:2005zs}. Vector fields might thus provide the actual and most 
genuine structure underlying any deformation-quantization.
But in order to consider such twist-deformations, an algebra of vector fields would have to be enhanced
to a Hopf-algebra. The actual question is, how this is possibly done. A very elegant solution to this
problem was provided by the authors of \cite{Majid:1998up}. But they already incorporated a physical
interpretation into their setup that we want to avoid here.
To any representation space we can formaly define an action of a Hopf-algebra of momenta. These can 
be joined to a left cross-product algebra that we devide in such a way, that we can lift it to an 
actual Hopf-algebra. In fact this construction provides a very clear and genuine structure that we 
further denote as a Hopf-algebra of vector fields. This Hopf-algebra is large enough to accomodate
any matrix representation of the universal enveloping algebra $U(g)$ as a subalgebra. This is the 
commutative limit that is well-known in physics and has to be fed into this setup. By twisting
the Hopf-algebra of vector fields we thus twist its subalgebra as well - but more general than
the generators of $U(g)$ could possibly do. In the mean time the twist is nothing else than the 
starproduct, that comes with the noncommutative associative space. We thus achieve several goals. 
Starproducts directly can be used as twists in order to obtain a quantization of the desired symmetry 
and in parallel we open the formalizm for most general quantizations and thus stay as close as possible 
to the actual requirements of physics.
The paper is organised as follows. In the first section we formulate the classical limit that we have to 
feed as input into our procedure. We take the opportunity to recall basic definitions and properties of 
required notions in order to be self-contained. In the following section we construct the Hopf-algebra of 
vector fields and the actual twists will be considered in the third section. We close with the basic 
example of a deformation of the two-dimensional representation of $U(sl_2)$. The exposition of the matter
orients itself to the textbooks \cite{Charipressley,Majid}.

\section{Representation of $U(\liealgebra)$ on $U(\X)$}

As outlined in the introduction, the deformation of a universal enveloping algebra $U(\liealgebra)$
of a Lie-algebra $\liealgebra$ and its accordingly deformed representation space $\X$ is actually 
independent of any representation theoretic issues, presupposing that the non-quantized limit 
exists and is well defined.

In this section we concretize this specific undeformed setup and in order to be self-contained we 
take the opportunity to recall basic definitions and properties of Lie-algebras and their 
representations.

It is our aim to represent $\liealgebra$ on a finite dimensional $\field$-linear vector space
$\X$. As fields $\field$ we consider complex or real numbers. Let us shortly recall the definition
of a Lie-algebra before we continue.

\begin{definition}[Lie-algebra]
	Let $\liealgebra$ be a $p$ - dimensional vector space over the field $\field$. The vector space
	$\liealgebra$ is called a \emph{Lie-algebra} if the there exists a bracket
	$$
		[\cdot, \cdot ]_\liealgebra : \liealgebra \times \liealgebra \to \liealgebra
	$$
	that holds the following properties:
	$$
		\begin{array}{lll}
			\forall g, h, k \in \liealgebra : & [g, h]_\liealgebra = - [h, g]_\liealgebra & 
				(\textrm{Antisymmetry}) \\
			& [g + h, k]_\liealgebra = [g, k]_\liealgebra + [h, k]_\liealgebra & (\textrm{Bilinearity}) \\
			& [g, [h, k]_\liealgebra ]_\liealgebra + [h, [k, g]_\liealgebra ]_\liealgebra
				+ [k, [g, h]_\liealgebra ]_\liealgebra = 0 & (\textrm{Jacobi-Identity})
		\end{array}
	$$
\end{definition}

As an element of the Lie-algebra $\liealgebra$, the bracket can be expressed as a linear combination in 
terms of basis elements $(g_a)_{a \in \{1, \ldots, p \}}$, i. e.

$$ \left[g_a, g_b\right]_\liealgebra = i \sum_{c = 1}^p f_{a b c} \; g_c, \;\;\;\; f_{a b c} \in \field . $$

Formally a representation of $\liealgebra$ on $\X$ is much more the representation of its universal enveloping algebra $U(\liealgebra)$ on $\X$, that we define as follows.

\begin{definition}[Universal Enveloping Algebra]
	Let $\liealgebra$ be a Lie-al\-ge\-bra over the field $\field$ with p-dimensional basis 
	$(g_a)_{a \in \{1, \ldots, p \}}$ and bracket $[\cdot , \cdot]_\liealgebra$. Then the 
	\emph{universal enveloping algebra} $U(\liealgebra)$ is defined to be the quotient of the 
	tensor algebra $T(\liealgebra)$ and the two-sided ideal $\mathcal{I}_\liealgebra \subset T(\liealgebra)$
	$$
		U(\liealgebra) = \frac{T(\liealgebra)}{\mathcal{I}_\liealgebra}.
	$$
	The two-sided ideal $\mathcal{I}_\liealgebra$ is generated by relations
	\beq
		\forall \; g_a, g_b \in \liealgebra : \; g_a \tensor g_b - g_b \tensor g_a - 
		i \sum_{c = 1}^p f_{a b c} \; g_c = 0 \label{liegenrel}
	\eeq
	For $\varphi(g_a), \omega(g_b) \in U(\liealgebra)$ the bracket $[\varphi(g_a), \omega(g_b)] := 
	\varphi(g_a) \tensor \omega(g_b) - \omega(g_a) \tensor \varphi(g_b)$ is called the \emph{commutator}.
\end{definition}

Before we continue to discuss our specific case let us also recall the definition of the representation
of an algebra on a $\field$-linear vector space.

\begin{definition}[Representation] Let $(\algebra, \product, \unit, +; \field)$ be an algebra over 
	the field $\field$ and let $(\V, +; \field)$ be a vector space. A \emph{left representation} of $\algebra$
	on $\V$ is a pair $(\rho, \V)$ consisting of a map
	\beqarr	
		\rho : \algebra \tensor \V & \longrightarrow & \V \nn \\
		a \tensor v & \mapsto & \rho(a \tensor v) = \rho_a(v) = a \laction v \nn
	\eeqarr
	such that for all $a \in \algebra$ the maps $\rho_a$ realize the algebra $\algebra$ within the 
	endomorphism of $\V$, i.e.
	\beqarr
		\forall \; a, b, \one \in \algebra, \; v \in \V & : & (a\cdot b) \laction v 
			= a \laction (b \laction v) \nn \\
		& & \one \laction v = v \nn
	\eeqarr
	The representation $\rho$ is also called a \emph{left action} "$\laction$".
\end{definition}
With this little preparation we understand that a representation $\rho$ of $U(\liealgebra)$
on the finite dimensional vector space $\X$ is more specifically defined in terms of a matrix
representation, i.e. for basis elements $g_a \in U(\liealgebra)$ and $x_i \in \X$ we obtain 
\beq
	\rho(g_a \tensor x_i)_j = (g_a \laction x_i)_j = \sum_{i = 1}^n (g_a)_{j\;i} x_i, \label{liematrixrep}
\eeq
where $(g_a)_{j\;i} \in GL(n, \field) \subset \textrm{Mat}(n, \field)$.
Moreover, the generating relations of $U(\liealgebra)$ have to be represented on $\X$ by
\beqarr
	\forall \; g_a, g_b \in \liealgebra 
		& : & (g_a \cdot g_b - g_b \cdot g_a - [g_a, g_b]_\liealgebra) \laction x_i \nn \\
		& & = g_a \laction ( g_b \laction x_i ) - g_b \laction ( g_a \laction x_i ) 
			- i \sum_{c = 1}^p f_{a b c} ( g_c \laction x_i ) = 0. \nn
\eeqarr
Here we replaced the tensor product "$\tensor$" by conventional multiplication "$\cdot$". In terms
of matrix representations (\ref{liematrixrep}) these relations then read
\beqarr
	\forall \; g_a, g_b \in \liealgebra 
		& : & \sum_{i = 1}^n(\sum_{j = 1}^n (g_a)_{k\;j} (g_b)_{j\;i} 
			- \sum_{j = 1}^n (g_b)_{k\;j} (g_a)_{j\;i} 
			- ([g_a, g_b]_\liealgebra)_{k\;i}) x_i \nn \\
		& & = \sum_{j = 1}^n (g_a)_{k\;j} (\sum_{i = 1}^n (g_b)_{j\;i} x_i)
			- \sum_{j = 1}^n (g_b)_{k\;j} (\sum_{i = 1}^n  (g_a)_{j\;i} x_i) \nn \\
		& & \;\;\;\; - i \sum_{c = 1}^p f_{a b c} \sum_{i = 1}^n (g_c)_{k\;i} x_i = 0
	\label{matrixrealization}
\eeqarr
Up to this point we consider the Lie-algebra $\liealgebra$ and the vector space $\X$ to be given 
and moreover that the representation $\rho$ exists and is well behaved. This setup represents the
actual input from outside that we require for our considerations. Of course we want more structure 
than that. For our purpose we have to enhance $\X$ to an algebra and thus extend $U(\liealgebra)$ to a 
Hopf-algebra. 

Enhancing $\X$ to an algebra is usually performed in several blends of one and the same 
idea: enhancing to the tensor algebra of $\X$ and then deviding by a suitable two-sided ideal. 
In order to get things straight, we first turn $\X$ into a Lie-algebra and then as well consider 
it as a universal enveloping algebra. 

We thus fix an $n$-dimensional basis for $\X$ to be $(x_i)_{i \in 1, 2, \ldots, n}$. Enhancing $\X$ to a Lie-algebra is easily performed by introducing a $\field$-bilinear bracket 

$$ [\cdot, \cdot ]\; : \; \X \times \X \longrightarrow \X. $$

The easiest choice for a bracket $[\cdot, \cdot]$, that satisfies the requirements of a Lie-algebra and 
later as well delivers the required commutative algebra of coordinates, is the vanishing bracket

$$ \forall \; x_i, x_j \in \X \; : \; [x_i, x_j] = 0. $$

We thus have turned $\X$ into a Lie-algebra. As we did for the Lie-algebra $\liealgebra$, we can now
consider the universal enveloping algebra $U(\X)$ of $\X$ and thus enhanced the vector space to a 
\emph{commutative} and associative algebra that is generated by relations
\beq
	\forall \; x_i, x_j \in U(\X): \; x_i \tensor x_j - x_j \tensor x_i = 0. \label{xrel}
\eeq
We once more replace the tensor product "$\tensor$" by a multiplication "$\cdot$". In order to transfer 
the action of $U(\liealgebra)$ on the vector space $\X$ to an action on the algebra $U(\X)$ we have to 
enhance $U(\liealgebra)$ to a Hopf-algebra by introducing a coproduct, counit and antipode by
\beq
	\forall \; g_a \in U(\liealgebra): \; \coproduct(g_a) = g_a \tensor \one + \one \tensor g_a
	, \;\; \counit(g_a) = 0
	, \;\; \antipode(g_a) = - g_a. \nn
\eeq
It is quickly verified that this definition of the Hopf-algebra $U(\liealgebra)$ satisfies all
axioms and requirements of a Hopf-algebra. The following definition then tells us how the 
representation $\rho$ on $\X$ is enhanced to that of $U(\X)$.
\begin{definition}
	Let $(\hopfalgebra, \product, \unit, \coproduct, \counit, \antipode; \field)$ be a Hopf-algebra over 
	the field $\field$. Let $(\algebra, \product, \unit, +; \field)$ be an algebra. The
	left representation of $\hopfalgebra$ on $\algebra$ is a left action that additionally satisfies
	\beqarr
		\forall \; h \in \hopfalgebra, \; a, b, \one \in \algebra 
			& : & h \laction (a \cdot b) = \sum (h_{(1)} \laction a) \cdot (h_{(2)} \laction b) \nn \\
			& & h \laction \one = \counit(h)
	\eeqarr
	with $\coproduct (h) = \sum h_{(1)} \tensor h_{(2)}$. The algebra $\algebra$ then becomes a
	\emph{left $\hopfalgebra$-module algebra}.
\end{definition}
Since the multiplication of $U(\X)$ is defined by the generating relations $ \forall \; x_i, x_j 
\in U(x_i): [x_i, x_j] = 0 $, we have to verify that the action of $U(\liealgebra)$ respects this, 
i.e. for $g_a \in U(\liealgebra)$
\beqarr
	& & g_a \laction (x_i \cdot x_j - x_j \cdot x_i) \nn \\
		& & \;\;\;\; = \coproduct(g_a) \laction (x_i \cdot x_j - x_j \cdot x_i) \nn \\
		& & \;\;\;\; = (g_a \laction x_i) x_j + x_i (g_a \laction x_j)
			- (g_a \laction x_i) x_j - x_i (g_a \laction x_j) \nn \\
		& & \;\;\;\; = (g_a \laction x_i) x_j - (g_a \laction x_i) x_j
			+ x_i (g_a \laction x_j) - x_i (g_a \laction x_j) = 0, \nn	
\eeqarr
since any $g_a \laction x_i \in U(\X)$ once more commutes with an $x_j \in U(\X)$. Thus the commutation
relations of $U(\X)$ have to be compatible with the coalgebra sector of $U(\liealgebra)$. We thus have
completed our setup that from now on is denoted by the \emph{commutative limit}. Note that we do \emph{not} 
enhance $U(\X)$ to a Hopf-algebra as well. In the next section we continue with basic constructions that 
pave the way to deformations of this setup.

\section{A Hopf-Algebra of Vector Fields $\mathfrak{W}(\Pi,\X)$}

In this section we construct the Hopf-algebra of vector fields $\mathfrak{W}(\Pi,\X)$ that we require 
for general deformations of $U(\liealgebra)$ and $U(\X)$. To this purpose we first introduce a 
Hopf-algebra of momenta $U(\Pi)$ that is represented as a left action on $U(\X)$. We continue with 
the construction of a \emph{left cross-product algebra} $U(\X) \leftcrossproduct \; U(\Pi)$ that we 
further devide in order to lift it to the Hopf-algebra of vector fields $\mathfrak{W}(\Pi,\X)$.
In the last subsection we further more define the left action of $\mathfrak{W}(\Pi,\X)$ on $U(\X)$.

\subsection{A Hopf-Algebra $U(\Pi)$ of Momenta}

We begin this section with one more Hopf-algebra $U(\Pi)$ that we loosely denote as the the 
\emph{algebra of momenta}. As long $U(\X)$ is actually considered to be an algebra of coordinates, 
$U(\Pi)$ can actually be considered to be nothing than that.

We introduce $U(\Pi)$ as a copy of $U(\X)$, with the exception that in contrast to $U(\X)$ it is
enhanced by coalgebra structure and an antipode. We thus understand $U(\Pi)$ to be generated by 
a $n$-dimensional basis $(\pi_i)_{i \in 1, 2, \ldots, n}$ with commutation relations
\beq 
	\pi_i \pi_j - \pi_j \pi_i = [\pi_i, \pi_j] = 0, \label{pirelations}
\eeq
and a primitive coalgebra structure for all $\pi_i \in U(\Pi)$ as well as a standard antipode 
\beq
	\coproduct(\pi_i) = \pi_i \tensor \one + \one \tensor \pi_i, \;\;\;\; 
	\counit(\pi_i) = 0, \;\;\;\; \antipode(\pi_i) = - \pi_i. \label{pico}
\eeq
We define the left action of $U(\Pi)$ on $U(\X)$ by 
\beq
	\forall \; \pi_i, \one \in U(\Pi) \; \wedge \; x_j, \one \in U(\X): 
		\pi_i \laction x_j = -i \delta_{i j}, \;
		\one \laction x_j = x_j, \;
		\pi_i \laction \one = \counit(\pi_i) \label{piaction}
\eeq
We could also have omitted the imaginary unit here, but since we are interested in physical applications,
we stick as close as possible to physical notions.
It is evident that (\ref{piaction}) is a well defined action, since the relations (\ref{pirelations}) 
are realized on $U(\X)$ by
\beqarr
	(\pi_i \pi_j - \pi_j \pi_i) \laction x_k 
		& = & \pi_i \laction (\pi_j \laction x_k) - \pi_j \laction (\pi_i \laction x_k) \nn \\
		& = & \pi_i \laction (-i \delta_{j k} \one) - \pi_j \laction (-i \delta_{i k} \one) = 0 
		\label{pipixaction}
\eeqarr
and in turn, $U(\Pi)$ respects the algebra relations (\ref{xrel}) of $U(\X)$ by means of the coalgebra 
structure (\ref{pico}) of $U(\Pi)$ by 
\beqarr
	\pi_i \laction (x_k x_l - x_l x_k) 
		& = & \coproduct(\pi_i) \laction (x_k x_l - x_l x_k) \nn \\
		& = & (\pi_i \laction x_k) x_l + x_k (\pi_i \laction x_l)
			- (\pi_i \laction x_l) x_k - x_l (\pi_i \laction x_k) \nn \\
		& = & - i\delta_{i k} x_l - i x_k \delta_{i l} + i \delta_{i l} x_k + i x_l \delta_{i k} = 0. \nn
\eeqarr

\subsection{The Left Cross-Product $U(\X) \leftcrossproduct \; U(\Pi)$}

Within the next step towards a Hopf-algebra of vector fields, we join the algebra $U(\X)$ and the 
Hopf-algebra $U(\Pi)$ to a single left cross-product algebra. Before we do so, we shortly recall its
definition-proposition, that can be found in the literature.

\begin{defproposition} 
	Let $\hopfalgebra$ be a Hopf-algebra and let $\algebra$ be a left $\hopfalgebra$-module
	algebra. Then there exists a left cross-product algebra $\algebra 
	\leftcrossproduct \hopfalgebra$ on $\algebra \tensor \hopfalgebra$ with the associative
	product
	$$
		\forall \; a, b \in \algebra, \; h, k \in \hopfalgebra: \; 
		(a \tensor h) \odot (b \tensor k) = \sum a (h_{(1)} \laction b) \tensor h_{(2)} k \nn
	$$
	and unit element $\one \tensor \one$.
\end{defproposition}

Thus for the algebraic relations of $U(\X) \leftcrossproduct \; U(\Pi)$, by the use of (\ref{pico}) and 
(\ref{piaction}), we obtain for $x_i \tensor \pi_r, x_j \tensor \pi_s \in U(\X) \tensor U(\Pi)$ 
\beqarr
	(x_i \tensor \pi_r) \odot (x_j \tensor \pi_s) 
		& = & x_i(\pi_r \laction x_j) \tensor \pi_s + x_i x_j \tensor \pi_r \pi_s \nn \\
		& = & - i \delta_{r j} x_i \tensor \pi_s + x_i x_j \tensor \pi_r \pi_s \nn
\eeqarr
In particular we compute that with $\coproduct(\one) = \one \tensor \one$ we obtain
\beqarr
	(x_i \tensor \one) \odot (x_j \tensor \one) & = & x_i x_j \tensor \one \nn \\
	(\one \tensor \pi_r) \odot (\one \tensor \pi_s) & = & \one \tensor \pi_r \pi_s, \nn
\eeqarr
such that $U(\X) \equiv U(\X) \tensor \one$ and $U(\Pi) \equiv \one \tensor U(\Pi)$ are contained as
subalgebras within $U(\X) \leftcrossproduct \; U(\Pi)$. We thus also find that
\beqarr
	[x_i \tensor \pi_r, x_j \tensor \pi_s]_\odot 
		& = & (x_i \tensor \pi_r) \odot (x_j \tensor \pi_s) - (x_j \tensor \pi_s) \odot (x_i \tensor \pi_r) \nn \\
		& = & - i \delta_{rj} x_i \tensor \pi_s + i \delta_{s i} x_j \tensor \pi_r. \nn
\eeqarr
Moreover, we find in particular that 
\beqarr
	\left[ x_i \tensor \pi_r, x_j \tensor \one \right]_\odot & = &  x_i (\pi_r \laction x_j) \tensor \one = 
		- i \delta_{rj} (x_i \tensor \one) \nn \\
	\left[ x_i \tensor \pi_r, \one \tensor \pi_s \right]_\odot & = & - (\pi_s \laction x_i) \tensor \pi_r = 
		i \delta_{s i}(\one \tensor \pi_r)  \nn \\
	\left[ \one \tensor \pi_r, x_j \tensor \one \right]_\odot & = & (\pi_r \laction x_j) \tensor \one = 
		- i \delta_{rj} (\one \tensor \one) \nn
\eeqarr
As $U(\X) \leftcrossproduct \; U(\Pi)$ provides the algebraic structure on $U(\X) \tensor U(\Pi)$,
that is a vector space, we can thus once more understand $U(\X) \leftcrossproduct \; U(\Pi)$ to be 
the tensor algebra $T(U(\X) \tensor U(\Pi))$ that is devided by a suitable two-sided ideal.
Making thus the identification
\beqarr
	\mathfrak{w}^0_{i r} \equiv x_i \tensor \pi_r, & & \mathfrak{w}^+_r \equiv \one \tensor \pi_r, \nn \\
	\mathfrak{w}^-_i \equiv x_i \tensor \one, & & \one \equiv \one \tensor \one, \nn
\eeqarr
we regard $\mathfrak{w^0}, \mathfrak{w^\pm}$ as the generators of $U(\X) \leftcrossproduct \; U(\Pi)$
that by relations 
\beqarr
	\left[\mathfrak{w}^0_{i r}, \mathfrak{w}^0_{j s} \right]_\odot 
		= - i \delta_{r j} \mathfrak{w}^0_{i s} + i \delta_{s i} \mathfrak{w}^0_{j r}, & &
		\left[\mathfrak{w}^+_r, \mathfrak{w}^-_j \right]_\odot = - i \delta_{rj} \one \nn \\	
	\left[\mathfrak{w}^0_{i r}, \mathfrak{w}^-_j \right]_\odot = - i \delta_{r j} \mathfrak{w}^-_i, & &
		\left[\mathfrak{w}^0_{i r}, \mathfrak{w}^+_s \right]_\odot = i \delta_{s i} \mathfrak{w}^+_r, \nn \\
	\left[\mathfrak{w}^+_r, \mathfrak{w}^+_s \right]_\odot = 0, & &
		\left[\mathfrak{w}^-_i, \mathfrak{w}^-_j \right]_\odot = 0, 
	\label{crossrelations}
\eeqarr
constitute the required two-sided ideal $\mathcal{I}_{\X, \Pi}$. We can thus set
$$
	U(\X) \leftcrossproduct \; U(\Pi) = \frac{T(U(\X) \tensor U(\Pi))}{\mathcal{I}_{\X, \Pi}},
$$
as for any universal enveloping algebra.

\subsection{The Hopf-algebra $\mathfrak{W}(\Pi,\X)$ of vector fields}

The Relations (\ref{crossrelations}) exhibit a nice structure of subalgebras within the cross-product algebra $U(\X) \leftcrossproduct \; U(\Pi)$, that already indicates into the desired direction of our purpose. However, since we would like to lift our construction to a Hopf-algebra, such that we can represent it once more on an algebra, we have to perform further modifications. The second relation of (\ref{crossrelations}) 
does not allow for a Hopf-algebra enhancement, since it would not confirm for the homomorphy property of
the coproduct. Moreover we do not really have a use for a coproduct on $\mathfrak{w}^-_i$, i.e. a coproduct on a coordinate. The authors of \cite{Majid:1998up} found an elegant way to deal with a similar issue by a 
specific bicross-product construcion. However, they had to introduce a physical interpretation as well that
we avoid here by the pursuing another direction.

We reach our goal by further deviding our algebra $U(\X) \leftcrossproduct \; U(\Pi)$ by relation
$$
	\mathfrak{w}^-_i = 0,
$$
such that we define our algebra of vector fields by 
$$
	\mathfrak{W}(\Pi,\X) = \frac{T(U(\X) \tensor U(\Pi))}{\mathcal{I}_\mathfrak{W}}.
$$
The two-sided ideal $\mathcal{I}_\mathfrak{W}$ is generated by relations
\beqarr
	\left[\mathfrak{w}^0_{i r}, \mathfrak{w}^0_{j s} \right]_\odot 
		= - i \delta_{r j} \mathfrak{w}^0_{i s} + i \delta_{s i} \mathfrak{w}^0_{j r}, & &
			\left[\mathfrak{w}^0_{i r}, \mathfrak{w}^+_s \right]_\odot = i \delta_{s i} \mathfrak{w}^+_r, \nn \\
	\left[\mathfrak{w}^+_r, \mathfrak{w}^+_s \right]_\odot = 0, & &	\mathfrak{w}^-_i = 0.
	\label{wrelations}
\eeqarr
We already see that this is very similar to the structure that we, for example, expect from a Poincar\'e-algebra. But it is much more general in its foundations. And we see how this applies to any desired 
setup based on the commutative limit we discussed above. It is easily checked that these relations 
induce a closed algebra, i.e. that the Jacobi-Identities
\beqarr
	\left[\left[\mathfrak{w}^0_{i r}, \mathfrak{w}^0_{j s} \right]_\odot, \mathfrak{w}^0_{k t} \right]_\odot +
	\left[\left[\mathfrak{w}^0_{j s}, \mathfrak{w}^0_{k t} \right]_\odot, \mathfrak{w}^0_{i r} \right]_\odot +
	\left[\left[\mathfrak{w}^0_{k t}, \mathfrak{w}^0_{i r} \right]_\odot, \mathfrak{w}^0_{j s} \right]_\odot 
	& = & 0 \nn \\
	\left[\left[\mathfrak{w}^0_{i r}, \mathfrak{w}^+_s \right]_\odot, \mathfrak{w}^0_{j t} \right]_\odot +
	\left[\left[\mathfrak{w}^+_s, \mathfrak{w}^0_{j t} \right]_\odot, \mathfrak{w}^0_{i r} \right]_\odot +
	\left[\left[\mathfrak{w}^0_{j t}, \mathfrak{w}^0_{i r} \right]_\odot, \mathfrak{w}^+_s \right]_\odot 
	& = & 0 \nn \\
	\left[\left[\mathfrak{w}^0_{i r}, \mathfrak{w}^+_s \right]_\odot, \mathfrak{w}^+_t \right]_\odot +
	\left[\left[\mathfrak{w}^+_s, \mathfrak{w}^+_t \right]_\odot, \mathfrak{w}^0_{i r} \right]_\odot +
	\left[\left[\mathfrak{w}^+_t, \mathfrak{w}^0_{i r} \right]_\odot, \mathfrak{w}^+_s \right]_\odot 
	& = & 0 \nn \\
	\left[\left[\mathfrak{w}^+_r, \mathfrak{w}^+_s \right]_\odot, \mathfrak{w}^+_t \right]_\odot +
	\left[\left[\mathfrak{w}^+_s, \mathfrak{w}^+_t \right]_\odot, \mathfrak{w}^+_r \right]_\odot +
	\left[\left[\mathfrak{w}^+_t, \mathfrak{w}^+_r \right]_\odot, \mathfrak{w}^+_s \right]_\odot 
	& = & 0 \nn
\eeqarr
are satisfied, as it should for an associative algebra of this kind.

We proceed by the following definition-proposition 
to enhance $\mathfrak{W}(\Pi,\X)$ to a Hopf-algebra.

\begin{defproposition} Let $\mathfrak{W}(\Pi,\X)$ be an algebra with the two-sided ideal $\mathcal{I}_\mathfrak{W}$, defined as above. Then $\mathfrak{W}(\Pi,\X)$ is a Hopf-algebra
with the following coproduct, counit and antipode
\beqarr
	\forall \; i, r \in 1, 2, \ldots n 
		& : & \coproduct(\mathfrak{w}^0_{i r}) 
			= \mathfrak{w}^0_{i r} \tensor \one + \one \tensor \mathfrak{w}^0_{i r}, \;\;\;\;
			\counit(\mathfrak{w}^0_{i r}) = 0, \nn \\
		& &	\antipode(\mathfrak{w}^0_{i r}) = - \mathfrak{w}^0_{i r}, \nn \\
		& & \coproduct(\mathfrak{w}^+_r) 
			= \mathfrak{w}^+_r \tensor \one + \one \tensor \mathfrak{w}^+_r, \;\;\;\;
			\counit(\mathfrak{w}^+_r) = 0, \nn \\
		& &	\antipode(\mathfrak{w}^+_r) = - \mathfrak{w}^+_r. \label{whopf}
\eeqarr
\end{defproposition}
\emph{Proof:} It is evident that the axioms of coassociativity 
$(\coproduct \tensor \id) \circ \coproduct = (\id \tensor \coproduct) \circ \coproduct$ and that of the 
counit $(\counit \tensor \id) \circ \coproduct = \id = (\id \tensor \counit) \circ \coproduct$ are 
satisfied for both, $\mathfrak{w}^0_{i r}$ and $\mathfrak{w}^+_r$. Moreover the antipode axiom 
$\product \circ (\antipode \tensor \id) \circ \coproduct = \unit \circ \counit = 
\product \circ (\id \tensor \antipode) \circ \coproduct$ is fulfilled as well. 
Here $\product$ is the multiplication within $\mathfrak{W}(\Pi,\X)$ and $\unit$ is the unit element, 
being the map
\beqarr
	\unit: \; \field & \longrightarrow & \mathfrak{W}(\Pi,\X) \nn \\
	\lambda & \mapsto & \lambda \cdot \one \nn
\eeqarr
Since it is an important issue here, we explicitly check on the homomorphy property of the coproduct.
Thus we check that
\beqarr
	\left[\coproduct(\mathfrak{w}^0_{i r}), \coproduct(\mathfrak{w}^0_{j s}) \right]_\odot 
		& = & \left[\mathfrak{w}^0_{i r}, \mathfrak{w}^0_{j s} \right]_\odot \tensor \one 
			+ \one \tensor \left[\mathfrak{w}^0_{i r}, \mathfrak{w}^0_{j s} \right]_\odot \nn \\
		& = & - i \delta_{r j} \coproduct(\mathfrak{w}^0_{i s}) 
			+ i \delta_{s i} \coproduct(\mathfrak{w}^0_{j r}) \nn
\eeqarr		
and 
\beqarr
	\left[\coproduct(\mathfrak{w}^0_{i r}), \coproduct(\mathfrak{w}^+_s) \right]_\odot 
	& = & \left[\mathfrak{w}^0_{i r}, \mathfrak{w}^+_s\right]_\odot \tensor \one 
		+ \one \tensor \left[\mathfrak{w}^0_{i r}, \mathfrak{w}^+_s\right]_\odot \nn \\
	& = & i \delta_{s i} \coproduct(\mathfrak{w}^+_r). \nn
\eeqarr
The same trivially holds for the counit. The antipode obviously is an anti-algebra homomorphism,
as it should, since
\beqarr
	- \left[\antipode(\mathfrak{w}^0_{i r}), \antipode(\mathfrak{w}^0_{j s}) \right]_\odot 
		& = & - i \delta_{r j} \antipode(\mathfrak{w}^0_{i s}) 
			+ i \delta_{s i} \antipode(\mathfrak{w}^0_{j r}) \nn \\
	- \left[\antipode(\mathfrak{w}^0_{i r}), \antipode(\mathfrak{w}^+_s) \right]_\odot 
		& = & i \delta_{s i} \antipode(\mathfrak{w}^+_r), \nn
\eeqarr
$\Box$

We are now prepared to consider representations of $\mathfrak{W}(\Pi,\X)$ on algebras.

\subsection{Representation of $\mathfrak{W}(\Pi,\X)$ on $U(\X)$}

It is our aim within this subsection to represent $\mathfrak{W}(\Pi,\X)$ on the algebra of 
coordinates $U(\X)$. Remember that we do not treat $U(\X)$ as a Hopf-algebra.
As vector fields, we introduce the \emph{left action} of $\mathfrak{w}^0_{i r}, 
\mathfrak{w}^+_s \in \mathfrak{W}(\Pi,\X)$ on $U(\X)$ by
\beqarr
	\mathfrak{w}^0_{i r} \laction x_j = x_i(\pi_r \laction x_j) = -i \delta_{r j} x_i, & &  
		\mathfrak{w}^0_{i r} \laction \one = \counit(\mathfrak{w}^0_{i r}), \nn \\
	\mathfrak{w}^+_r \laction x_j = \pi_r \laction x_j = -i \delta_{r j} \one, & &
	\mathfrak{w}^+_r \laction \one = \counit(\mathfrak{w}^+_r).
	\label{wrep}
\eeqarr
We have thus to verify that the Hopf-algebra of vector fields $\mathfrak{W}(\Pi,\X)$ is realized
as vector space endomorphisms on $U(\X)$. In particular this means that the first two relations of (\ref{wrelations}) have to be realized by means of (\ref{wrep}), i.e. we obtain
\beqarr
	& & \left(\left[\mathfrak{w}^0_{i r}, \mathfrak{w}^0_{j s} \right]_\odot 
		+ i \delta_{r j} \mathfrak{w}^0_{i s} - i \delta_{s i} \mathfrak{w}^0_{j r} \right) \laction x_k \nn \\
	& & \;\;\;\;\;\; = \mathfrak{w}^0_{i r} \laction \left( \mathfrak{w}^0_{j s} \laction x_k \right)
		- \mathfrak{w}^0_{j s} \laction \left( \mathfrak{w}^0_{i r} \laction x_k \right)
		+ i \delta_{r j} \left( \mathfrak{w}^0_{i s} \laction x_k \right)
		- i \delta_{s i} \left( \mathfrak{w}^0_{j r} \laction x_k \right) \nn \\
	& & \;\;\;\;\;\; = \mathfrak{w}^0_{i r} \laction \left( - i \delta_{s k} x_j \right)
		- \mathfrak{w}^0_{j s} \laction \left( - i \delta_{r k} x_i \right)
		+ i \delta_{r j} \left(- i \delta_{s k} x_i \right)
		- i \delta_{s i} \left(- i \delta_{r k} x_j \right) \nn \\
	& & \;\;\;\;\;\; = - \delta_{s k}	\delta_{r j} x_i + \delta_{r k}	\delta_{s i} x_j 
		+ \delta_{j r}	\delta_{s k} x_i - \delta_{s i}	\delta_{r k} x_j = 0 \nn
\eeqarr
and 
\beqarr		
	& & \left(\left[\mathfrak{w}^0_{i r}, \mathfrak{w}^+_s \right]_\odot 
		- i \delta_{s i} \mathfrak{w}^+_r \right) \laction x_j \nn \\
	& & \;\;\;\;\;\; = \mathfrak{w}^0_{i r} \laction \left(\mathfrak{w}^+_s \laction x_j \right)
		- \mathfrak{w}^+_s \laction \left(\mathfrak{w}^0_{i r} \laction x_j \right)
		- i \delta_{s i} \left( \mathfrak{w}^+_r \laction x_j \right) \nn \\
	& & \;\;\;\;\;\; = \mathfrak{w}^0_{i r} \laction \left(- i \delta_{s j} \one \right)
		- \mathfrak{w}^+_s \laction \left( - i \delta_{r j} x_i \right)
		- i \delta_{s i} \left( - i \delta_{r j} \one \right) \nn \\
	& & \;\;\;\;\;\; = \delta_{r j}\delta_{s i} - \delta_{s i}\delta_{r j} = 0. \nn
\eeqarr
The third relation is already represented on $U(\X)$ given by (\ref{pipixaction}).
We further more have to check whether the representation (\ref{wrep}) respects the algebra relations
(\ref{xrel}) of $U(\X)$, i.e. we have
\beqarr
	& & \mathfrak{w}^0_{i r} \laction (x_j x_k - x_k x_j) 
		= \coproduct(\mathfrak{w}^0_{i r}) \laction (x_j x_k - x_k x_j) \nn \\
	& & \;\;\;\;\;\; = (\mathfrak{w}^0_{i r} \laction x_j) x_k + x_j ( \mathfrak{w}^0_{i r} \laction x_k )
		- (\mathfrak{w}^0_{i r}\laction x_k ) x_j - x_k ( \mathfrak{w}^0_{i r} \laction x_j ) \nn \\
	& & \;\;\;\;\;\; = (- i \delta_{r j} x_i ) x_k + x_j (- i \delta_{r k} x_i )
		- (- i \delta_{r k} x_i ) x_j - x_k ( - i \delta_{r j} x_i ) = 0 \nn
\eeqarr
and
\beqarr
	& & \mathfrak{w}^+_r \laction (x_j x_k - x_k x_j) 
		= \coproduct(\mathfrak{w}^+_r) \laction (x_j x_k - x_k x_j) \nn \\
	& & \;\;\;\;\;\; = (\mathfrak{w}^+_r \laction x_j) x_k + x_j(\mathfrak{w}^+_r \laction x_k)
		- (\mathfrak{w}^+_r \laction x_k )x_j - x_k (\mathfrak{w}^+_r \laction x_j) \nn \\
	& & \;\;\;\;\;\; = ( - i\delta_{r j} ) x_k + x_j( - i \delta_{r k} )
		- ( - i \delta_{r k} )x_j - x_k ( - i \delta_{r j} ) = 0. \nn	
\eeqarr
We thus made all necessary preparations to attack the actual interesting step in the next section.

\section{Representation of $U(\liealgebra)$ in $\mathfrak{W}(\Pi,\X)$}

In this section we map $U(\liealgebra)$ as a subalgebra within $\mathfrak{W}(\Pi,\X)$ by means of
its matrix representation (\ref{liematrixrep}) and the Hopf-algebra homomorphism
\beqarr
	\rho : \; U(\liealgebra) & \longrightarrow & \mathfrak{W}(\Pi,\X) \nn \\
	g_a & \mapsto & i (g_a)_{r i} \mathfrak{w}^0_{i r}. \nn
\eeqarr
We verify that the generating relations (\ref{liegenrel}) of $U(\liealgebra)$ are realized in terms of 
(\ref{wrelations}). In particular, we obtain for basis elements $g_a, g_b \in U(\liealgebra)$
\beqarr
	\left[g_a, g_b \right]_\odot
		& = & \left[(g_a)_{r i}\mathfrak{w}^0_{i r}, (g_b)_{s j} \mathfrak{w}^0_{j s}\right]_\odot
		=	(g_a)_{r i} (g_b)_{s j} \left[\mathfrak{w}^0_{i r}, \mathfrak{w}^0_{j s}\right]_\odot \nn \\
	& = &	(g_a)_{r i} (g_b)_{s j} 
		\left( - i \delta_{r j} \mathfrak{w}^0_{i s} + i \delta_{s i} \mathfrak{w}^0_{j r} \right) \nn \\
	&	= & - i (g_b)_{s k}	(g_a)_{k i}	\mathfrak{w}^0_{i s} + i (g_a)_{r k} (g_b)_{k j} \mathfrak{w}^0_{j r} \nn \\
	& = & i ( (g_a)_{s k} (g_b)_{k i} - (g_b)_{s k}	(g_a)_{k i}) \mathfrak{w}^0_{i s}
		= i f_{a b c} i (g_c)_{s i} \mathfrak{w}^0_{i s} = i f_{a b c} g_c \nn
\eeqarr
Here we use summation convention for any pair of equal indices. The Hopf structure (\ref{whopf}) $\mathfrak{W}(\Pi,\X)$ corresponds to that of $U(\liealgebra)$, i. e.
\beqarr
	\coproduct(g_a) & = & \coproduct(i (g_a)_{r i} \mathfrak{w}^0_{i r})
		= i (g_a)_{r i} \coproduct(\mathfrak{w}^0_{i r}) 
		= i (g_a)_{r i} \left( \mathfrak{w}^0_{i r} \tensor \one + \one \tensor \mathfrak{w}^0_{i r} \right) \nn \\
	& = & g_a \tensor \one + \one \tensor g_a \nn \\
	\counit(g_a) & = & \counit(i (g_a)_{r i} \mathfrak{w}^0_{i r}) 
		= i (g_a)_{r i} \counit(\mathfrak{w}^0_{i r}) = 0 \nn \\
	\antipode(g_a) & = & \antipode(i (g_a)_{r i} \mathfrak{w}^0_{i r}) 
		= i (g_a)_{r i} \antipode(\mathfrak{w}^0_{i r}) = - i (g_a)_{r i} \mathfrak{w}^0_{i r} = - g_a \nn 
\eeqarr
We verify that the representation of $U(\liealgebra)$ in $\mathfrak{W}(\Pi,\X)$ also accomodates the correct
representation on $U(\X)$. The representation of $\mathfrak{W}(\Pi,\X)$ on $U(\X)$ implies that
\beqarr
	\left( g_a \laction x_i \right)_k 
		& = & \left(\left( i (g_a)_{s j} \mathfrak{w}^0_{j s} \right) \laction x_i \right)_k
		= \left( i (g_a)_{s j} \left( \mathfrak{w}^0_{j s} \laction x_i \right) \right)_k \nn \\
	& = & \left(i (g_a)_{s j} \left( - i \delta_{s i} x_j \right) \right)_k 
		= \left((g_a)_{i j} x_j \right)_k = (g_a)_{k j} x_j \nn
\eeqarr
This neatly corresponds to the matrix representation (\ref{liematrixrep}). We obtain double applications
of the represented generators of $U(\liealgebra)$ according to
\beqarr
	\left( \left(g_a g_b \right) \laction x_i \right)_k 
	& = & \left( i (g_b)_{s j} \mathfrak{w}^0_{j s} 
		\laction (i (g_a)_{r l} \mathfrak{w}^0_{l r} \laction x_i )\right)_k \nn \\
	& = & \left(- (g_b)_{s j} (g_a)_{r l} \mathfrak{w}^0_{j s} 
		\laction \left(- i \delta_{i r} x_l \right)\right)_k \nn \\
	& = & \left(- (g_b)_{s j} (g_a)_{r l} ( -i \delta_{i r}) 
		\left( \mathfrak{w}^0_{j s} \laction x_l \right) \right)_k \nn \\
	& = & \left(- (g_b)_{s j} (g_a)_{r l} ( -i \delta_{i r}) (- i \delta_{l s}) x_j \right)_k \nn \\
	& = & \left( (g_a)_{i l} (g_b)_{l j} x_j \right)_k = (g_a)_{k l} (g_b)_{l j} x_j \nn
\eeqarr
Note that the formal reversal of the order of generators $\mathfrak{w}^0$ is only applied to get indices 
straight. The actual order of application of generators remains unchanged as one can see from the last 
equation. We once more verify that this actually realizes the generating relations (\ref{liegenrel}) of $U(\liealgebra)$ on $U(\X)$ via matrix representation according to (\ref{matrixrealization}), i.e.
\beqarr
	\left( \left(g_a g_b - g_b g_a\right) \laction x_i \right)_k 
		& = & \left( \left( (g_a)_{i l} (g_b)_{l j} - (g_b)_{i l} (g_a)_{l j} \right) x_j \right)_k \nn \\
	& = & \left( \left( i f_{a b c} (g_c)_{i j} \right) x_j \right)_k 
		=  \left( i f_{a b c} \left( i (g_c)_{s j} \mathfrak{w}^0_{j s} \right) \laction x_i \right)_k \nn \\
	& = & \left( i f_{a b c} ( g_c \laction x_i) \right)_k \nn
\eeqarr
Through the coproduct in $\mathfrak{W}(\Pi,\X)$ it is clear that our realization of $U(\liealgebra)$ in
$\mathfrak{W}(\Pi,\X)$ respects the generating relations of $U(\X)$. We thus have received a left action
of the Hopf-algebra $U(\liealgebra)$ on $U(\X)$ via its matrix representation within $\mathfrak{W}(\Pi,\X)$. 
We can now proceed to twist $\mathfrak{W}(\Pi,\X)$ and thus to most generally twist its subalgebra 
$U(\liealgebra)$ as well.

\section{Twisting}

In order to obtain deformations $\mathfrak{W}(\Pi,\X)$, we introduce twists in this section. To this purpose 
we recall some basic properties of twists. Since we want to consider the twists of vector fields to be
starproducts of associative algebras of coordinates $U(\X)$ at the same time, it is our intend to clearify
that the definition of twists incorporates this demand. We then proceed and give some examples of twists
for $\mathfrak{W}(\Pi,\X)$ that we apply to a two-dimensional representation of $U(sl_2)$ in the next 
section. For this section we recommend \cite{Charipressley} as a textbook for reference. We begin by recalling 
the definition of a twist.

\begin{definition} Let $\left(\hopfalgebra, \product, \unit, \coproduct, \counit, \antipode ; 
\field \right)$ be a Hopf-algebra over the field $\field$. Then an invertible object 
$\twist \in \hopfalgebra \tensor \hopfalgebra$ is called a \emph{twist}, if the following 
two conditions hold
\beqarr
	\twist_{1 2}\left(\coproduct \tensor \id \right)(\twist) 
		& = & \twist_{2 3} \left(\id \tensor \coproduct \right)(\twist) \label{twistdefco} \\
	\left(\counit \tensor \id \right)(\twist) & = 1 = & \left( \id \tensor \counit \right)(\twist). 
		\label{twistdefcoun} 
\eeqarr
For $\twist = \sum \twist^{(1)} \tensor \twist^{(2)}$ the objects $\twist_{1 2}$ and $\twist_{2 3}$ 
are defined by
\beqarr
	\twist_{1 2} & = & \sum \twist^{(1)} \tensor \twist^{(2)} \tensor \one \nn \\
	\twist_{2 3} & = & \sum \one \tensor \twist^{(1)} \tensor \twist^{(2)}. \nn
\eeqarr
\end{definition}

Using this definition, we can now recall the required proposition stating how a twist is used to 
deform the corresponding Hopf-algebra.

\begin{proposition} Let $\left(\hopfalgebra, \product, \unit, \coproduct, \counit, \antipode; 
	\field \right)$ be a Hopf-algebra and let furthermore the objects $\eta, \eta^{-1} \in \hopfalgebra$ 
	be given by 
	\beqarr
		\eta & = & \product \left(\id \tensor \antipode \right)(\twist) \nn \\
		\eta^{-1} & = & \product \left(\antipode \tensor \id \right) (\twist).\nn 
	\eeqarr
	Then $\left(\hopfalgebra, \product, \unit, \coproduct_\twist, \counit, \antipode_\twist; \field \right)$
	with 
	\beqarr
		\coproduct_\twist (h) & = & \twist \coproduct (h) \twist^{-1} \nn \\
		\antipode_\twist (h) & = & \eta \antipode(h)\eta^{-1}\nn
	\eeqarr
	and $h\in \hopfalgebra$ is the Hopf-algebra $\hopfalgebra_\twist$ that is called the 
	\emph{twist of} $\hopfalgebra$. 
\end{proposition}

Note that the Hopf-algebra $\hopfalgebra$ not necessarily has to be cocommutative. We further elucidate
some consequences and properties of the defined twist before we come to specific examples for 
$\mathfrak{W}(\Pi,\X)$. If the Hopf-algebra $\hopfalgebra$ is represented on $U(\X)$ by a left 
action, then the generating relations (\ref{xrel}) of $U(\X)$ are preserved under the action of 
$\hopfalgebra$, i.e. for $h \in \hopfalgebra$ we have
\beqarr
	x_i x_j - x_j x_i = 0 & \Rightarrow & h \laction \left( x_i x_j - x_j x_i \right) \nn \\
	& & = \coproduct(h) \laction \left( x_i x_j - x_j x_i \right) \nn \\
	& & = \sum (h_{(1)} \laction x_i)( h_{(2)} \laction x_j) - ( h_{(1)} \laction x_j )( h_{(2)} \laction x_i) = 0. \nn
\eeqarr
Within the representation of $\hopfalgebra$ on $U(\X)$ we can consider a twist $\twist \in \hopfalgebra \tensor \hopfalgebra$ to deform the product $\product$ of $U(\X)$ to a noncommutative product $\product_\twist$ by
$$
	\product_\twist(x_i, x_j) = 
	x_i *_\twist x_j =  \twist^{-1} \laction (x_i \cdot x_j) 
	= \product\left((\twist^{-1\;\;(1)} \laction x_i),(\twist^{-1\;\;(2)} \laction x_j) \right).
$$
This implies new generating relations for a deformation of $U(\X)$, that we further denote 
by $U(\X_\twist)$, being
\beq
	x_i *_\twist x_j - x_j *_\twist x_i - \left[x_i \stackrel{*_\twist}{,} x_j \right]= 0,
	\label{xtwistrel}
\eeq
where the commutator $\left[x_i \stackrel{*_\twist}{,} x_j \right]$ has to be replaced by a 
corresponding right hand side. This nonvanishing commutator reflects the noncocommutativity 
of the twisted coproduct $\coproduct_\twist$ in $\hopfalgebra_\twist$. The defining relations 
(\ref{twistdefco}) and (\ref{twistdefcoun}) of the twist $\twist$ thereby ensure that the 
axiom of coassociativity and the counit axiom of the coproduct $\coproduct_\twist$ are 
satisfied, i.e. that
\beqarr	
	(\coproduct_\twist \tensor \id) \circ \coproduct_\twist 
		& = & (\id \tensor \coproduct_\twist) \circ \coproduct_\twist \nn \\
	(\counit \tensor \id) \circ \coproduct_\twist 
		& = & (\id \tensor \counit) \circ \coproduct_\twist \nn
\eeqarr
Covariance of the generating relations (\ref{xtwistrel}) of $U(\X_\twist)$ under 
the action of $\hopfalgebra_\twist$ is then given by
\beqarr 
	& & h \laction \left( x_i *_\twist x_j - x_j *_\twist x_i 
		- \left[x_i \stackrel{*_\twist}{,} x_j \right] \right) \nn \\
	& & \;\;\;\;\;\; = h \laction (\twist^{-1} \laction (x_i \cdot x_j))
		- h \laction (\twist^{-1} \laction (x_j \cdot x_i))
		- h \laction \left[x_i \stackrel{*_\twist}{,} x_j \right] \nn \\
	& & \;\;\;\;\;\; = \twist^{-1} \laction \left(\twist \coproduct(h) \twist^{-1} \right) \laction 
		( x_i \cdot x_j ) \nn \\
	& & \;\;\;\;\;\;\;\;\;\; - \twist^{-1} \laction \left(\twist \coproduct(h) \twist^{-1} \right) 
			\laction (x_j \cdot x_i) - h \laction \left[x_i \stackrel{*_\twist}{,} x_j \right] \nn \\
	& & \;\;\;\;\;\; =	\twist^{-1} \laction \left(\coproduct_\twist(h) \laction (x_i \cdot x_j)   
		- \coproduct_\twist(h) \laction (x_j \cdot x_i)\right)	 
		- h \laction \left[x_i \stackrel{*_\twist}{,} x_j \right] \nn
\eeqarr
Thus transformation and deformation commute, such that noncommutativity of $U(\X_\twist)$ is preserved under 
the left action of $\hopfalgebra_\twist$. Coassociativity of $\coproduct_\twist$ implies the associativity 
of the starproduct $*_\twist$, i.e. we have 
\beqarr
	\twist \laction \left( h \laction \left( x_i *_\twist (x_j *_\twist x_i) \right) \right)
	& = & \left(\id \tensor \coproduct_\twist \right) \circ \coproduct_\twist(h) 
		\laction \left( x_i \cdot x_j \cdot x_i) \right) \nn \\
	& = & \left(\coproduct_\twist \tensor \id \right) \circ \coproduct_\twist(h) 
		\laction \left( x_i \cdot x_j \cdot x_i) \right) \nn \\	
	& = & \twist \laction \left( h \laction \left( (x_i *_\twist x_j) *_\twist x_i \right) \right) \nn 
\eeqarr
In the following we consider specific twistings of $\mathfrak{W}(\Pi,\X)$. It is our intend to 
merely outline the application of the formalism. We thus stick to some simple but nontrivial and 
genuine examples. We encourage the reader to derive more sophisticated twists for his very own 
purpose and use the following consideration as an examplary guiding line. Our first example is 
given by
\beq
	\twist_\theta := e^{\frac{i}{2} \theta_{r s}\mathfrak{w}^+_r \tensor \mathfrak{w}^+_s}, 
		\;\;\;\; \theta_{r s} = - \theta_{s r} \in \R \label{thetatwist}
\eeq
Since $\counit(\mathfrak{w}^+_r) = 0$ relation (\ref{twistdefcoun}) is satisfied. Relation (\ref{twistdefco}) is
satisfied as well since 
\beqarr
	& & e^{\frac{i}{2} \theta_{r s}\mathfrak{w}^+_r \tensor \mathfrak{w}^+_s \tensor \one} 
	(\coproduct \tensor \id)(e^{\frac{i}{2} \theta_{r s}\mathfrak{w}^+_r \tensor \mathfrak{w}^+_s}) \nn \\
	& & \;\;\;\;\;\;\;\;\; = e^{\frac{i}{2} \theta_{r s} (\mathfrak{w}^+_r \tensor \mathfrak{w}^+_s \tensor \one
		+ \mathfrak{w}^+_r \tensor \one \tensor \mathfrak{w}^+_s
		+ \one \tensor \mathfrak{w}^+_r \tensor \mathfrak{w}^+_s)} \nn \\
	& & \;\;\;\;\;\;\;\;\; = e^{\frac{i}{2} \theta_{r s}\one \tensor \mathfrak{w}^+_r \tensor \mathfrak{w}^+_s} 
	(\id \tensor \coproduct)(e^{\frac{i}{2} \theta_{r s}\mathfrak{w}^+_r \tensor \mathfrak{w}^+_s}), \nn
\eeqarr
due to the vanishing commutator $[\mathfrak{w}^+_r, \mathfrak{w}^+_s] = 0$. In general these computations are 
performed using the Baker-Campbell-Hausdorff formula
$$
	e^A \; e^B = e^{A + B + \frac{1}{2}[A,B] + \frac{1}{12}([A, [A, B]] - [B, [A, B]]) 
		+ \frac{1}{48}([A, [B, [B, A]]] - [B, [A, [A, B]]]) + \ldots }.
$$
Using the formula
\beq
	e^A \; B \; e^{-A} = \sum_{n = 0}^\infty \frac{1}{n !}
		\left[A, \left[A, \left[A, \ldots \left[A, B \right] \right] \right] \right]
	\label{conformula}
\eeq
we can now compute the twisted coproducts of $\mathfrak{w}^+_r$ and $\mathfrak{w}^0_{i r}$ to be
\beqarr
	\coproduct_\twist(\mathfrak{w}^+_r) & = & \mathfrak{w}^+_r \tensor \one + \one \tensor \mathfrak{w}^+_r \nn \\
	\coproduct_\twist(\mathfrak{w}^0_{i r}) 
		& = & \mathfrak{w}^0_{i r} \tensor \one + \one \tensor \mathfrak{w}^0_{i r} \nn \\
	& & - \frac{1}{2}\theta_{i s}\left(\mathfrak{w}^+_s \tensor \mathfrak{w}^+_r - \mathfrak{w}^+_r 
		\tensor \mathfrak{w}^+_s \right). \nn 
\eeqarr
These of course correspond to the results of \cite{Chaichian:2004za}, but now this twist can be applied
to any representation of a universal enveloping algebra $U(\liealgebra)$. We obtain the generating relations 
of $U(\X_{\twist_\theta})$ by
\beqarr
	& & x_i *_{\twist_\theta} x_j - x_j *_{\twist_\theta} x_i \nn \\
	& & \;\;\;\;\;\;\;\;\; 
		= e^{-\frac{i}{2} \theta_{r s}\mathfrak{w}^+_r \tensor \mathfrak{w}^+_s} \laction (x_i x_j)
		- e^{-\frac{i}{2} \theta_{r s}\mathfrak{w}^+_r \tensor \mathfrak{w}^+_s} \laction (x_j x_i) \nn \\
	& & \;\;\;\;\;\;\;\;\; 
		= (1 -\frac{i}{2} \theta_{r s}\mathfrak{w}^+_r \tensor \mathfrak{w}^+_s)\laction (x_i x_j)
		- (1 -\frac{i}{2} \theta_{r s}\mathfrak{w}^+_r \tensor \mathfrak{w}^+_s)\laction (x_j x_i) \nn \\
	& & \;\;\;\;\;\;\;\;\; 
		= x_i x_j -\frac{i}{2} \theta_{r s} (-i \delta_{r i})(-i \delta_{s j})    
		- x_j x_i + \frac{i}{2} \theta_{r s} (-i \delta_{r j})(-i \delta_{s i}) \nn \\
	& & \;\;\;\;\;\;\;\;\; = i \; \theta_{i j} \nn
\eeqarr
We come now to a more genuine example taken from \cite{Jambor:2004kc}. We introduce the twist
\beq
	\twist_h := e^{i \; h \; \mathfrak{w}^0_{1 1} \tensor \mathfrak{w}^+_2}. \label{htwist}
\eeq
The generators $\mathfrak{w}^0_{1 1}$ and $\mathfrak{w}^+_2$ commute according to (\ref{wrelations}), 
i.e. $[\mathfrak{w}^0_{1 1}, \mathfrak{w}^+_2] = 0$. Relation (\ref{twistdefcoun}) once more is trivially
satisfied. We check for (\ref{twistdefco}), i.e.
\beqarr
	& & e^{i \; h \; \mathfrak{w}^0_{1 1} \tensor \mathfrak{w}^+_2 \tensor \one} 
		\left( \coproduct \tensor \id \right)( e^{i \; h \; \mathfrak{w}^0_{1 1} \tensor \mathfrak{w}^+_2} ) \nn \\
	& & \;\;\;\;\;\;\;\; =  e^{i \; h \; ( \mathfrak{w}^0_{1 1} \tensor \mathfrak{w}^+_2 \tensor \one 
		+ \mathfrak{w}^0_{1 1} \tensor \one \tensor \mathfrak{w}^+_2
		+ \one \tensor \mathfrak{w}^0_{1 1} \tensor \mathfrak{w}^+_2 ) } \nn \\
	& & \;\;\;\;\;\;\;\; = e^{i \; h \; \one \tensor \mathfrak{w}^0_{1 1} \tensor \mathfrak{w}^+_2}
			\left( \id \tensor \coproduct \right)( e^{i \; h \; \mathfrak{w}^0_{1 1} \tensor \mathfrak{w}^+_2} ). \nn
\eeqarr
We once more derive the twisted coproducts using formula (\ref{conformula}). The coproducts of 
$\mathfrak{w}^+_s$ remain undeformed for $s \not = 1$. For the coproduct of $\mathfrak{w}^+_1$, we obtain
\beq
	\coproduct_{\twist_h}(\mathfrak{w}^+_1) = \mathfrak{w}^+_1 \tensor \one + \one \tensor \mathfrak{w}^+_1
	+ \mathfrak{w}^+_1 \tensor \left(e^{- h\mathfrak{w}^+_2} - 1 \right)
\eeq
The twisted coproduct of $\mathfrak{w}^0_{i r}$ also remains undeformed apart from four specific cases, that are
\beqarr
	r \not = 1 & : & \coproduct_{\twist_h}(\mathfrak{w}^0_{1 r}) 
		= \mathfrak{w}^0_{1 r} \tensor \one + \one \tensor \mathfrak{w}^0_{1 r}
			+ \mathfrak{w}^0_{1 r} \tensor \left(e^{+ h\mathfrak{w}^+_2} - 1 \right) \nn \\
	i \not = 1, 2 & : & \coproduct_{\twist_h}(\mathfrak{w}^0_{i 1})
		= \mathfrak{w}^0_{i 1} \tensor \one + \one \tensor \mathfrak{w}^0_{i 1}
			+ \mathfrak{w}^0_{i 1} \tensor \left(e^{- h\mathfrak{w}^+_2} - 1 \right) \nn \\
	i = 2, r = 1 & : & \coproduct_{\twist_h}(\mathfrak{w}^0_{2 1})
		= \mathfrak{w}^0_{2 1} \tensor \one + \one \tensor \mathfrak{w}^0_{2 1}
			+ h \; \mathfrak{w}^0_{1 1} \tensor \mathfrak{w}^+_1 \nn \\
	& & + \mathfrak{w}^0_{2 1} \tensor \left(e^{- h \; \mathfrak{w}^+_2} - 1 \right) \nn \\
	r \not = 1 & : & \coproduct_{\twist_h}(\mathfrak{w}^0_{2 r})
		= \mathfrak{w}^0_{2 r} \tensor \one + \one \tensor \mathfrak{w}^0_{2 r}
			+ h \; \mathfrak{w}^0_{1 1} \tensor \mathfrak{w}^+_r. \nn
\eeqarr
The generating relations of $U(\X_{\twist_h})$ are then given by 
\beqarr
	& & x_i *_{\twist_h} x_j - x_j *_{\twist_h} x_i \nn \\
	& & \;\;\;\;\;\;\;\;	= e^{- i \; h \; \mathfrak{w}^0_{1 1} \tensor \mathfrak{w}^+_2} \laction (x_i x_j)
		- e^{- i \; h \; \mathfrak{w}^0_{1 1} \tensor \mathfrak{w}^+_2} \laction (x_j x_i) \nn \\
	& & \;\;\;\;\;\;\;\;	=	(1 - i \; h \; \mathfrak{w}^0_{1 1} \tensor \mathfrak{w}^+_2) \laction (x_i x_j)
		- (1 - i \; h \; \mathfrak{w}^0_{1 1} \tensor \mathfrak{w}^+_2) \laction (x_j x_i) \nn \\
	& & \;\;\;\;\;\;\;\;	=	x_i x_j  + i \; h \delta_{i 1} \delta_{j 2} x_1
		- x_j x_i - i \; h \delta_{j 1} \delta_{i 2} x_1 \nn \\
	& & \;\;\;\;\;\;\;\;	= i \; h ( \delta_{i 1} \delta_{j 2} - \delta_{j 1} \delta_{i 2}) x_1. \nn
\eeqarr

We thus see in this final example how the introduced formalism of vector fields $\mathfrak{W}(\Pi,\X)$ 
unfolds its impact. The twist $\twist_h$ cannot be expressed in terms of generators of $U(\liealgebra)$ 
but through the representation of  $U(\liealgebra)$ in $\mathfrak{W}(\Pi,\X)$ we now, nevertheless, use 
it to twist its coproduct and thus obtain the desired deformation of the symmetry algebra. This is sketched 
in the next section at the example of $U(sl_2)$.

\section{Deformation of a two-dimensional Representation of $U(sl_2)$}

In this section we shortly consider the two-dimensional representation of $U(sl_2)$ that we want to twist 
by means of (\ref{htwist}). To this purpose we directly consider the corresponding matrix representation
of $U(sl_2)$ given in terms of Pauli-matrices and a canonical basis for the representation space. 
The Hopf-algebra of $U(sl_2)$ can thus be considered to be generated by the basis 
$(\sigma_i)_{i \in 1,2,3}$ with the Hopf-structure 
$$
	\coproduct(\sigma_i) = \sigma_i \tensor \one + 
		\one \tensor \sigma_i, \;\;\; \counit(\sigma_i) = 0, \;\;\; \antipode(\sigma_i) = - \sigma_i
$$
In the two-dimensional representation we then identify with the well-known Pauli-matrices:
$$ 
	\sigma_1 = \left( \begin{array}{cc}
															0 & 1 \\
															1	& 0 
														\end{array}
 											\right), \;\;
 	\sigma_2 = \left( \begin{array}{cc}
															0 & -i \\
															i	& 0 
														\end{array}
 											\right), \;\;
	\sigma_3 = \left( \begin{array}{cc}
															1 & 0 \\
															0	& -1 
														\end{array}
 											\right)										
$$
Making the identification 
$$
	x_1 = \left( \begin{array}{c}
												1 \\
												0	
							\end{array}
 			\right), \;\;\;\;\;\;
 	x_2 = \left( \begin{array}{c}
												0 \\
												1	
							\end{array},
 			\right)
$$
we obtain the explicit left action of the two-dimensional representation of $U(sl_2)$ by 
\beqarr
	& & \sigma_1 \laction x_1 = x_2, \;\; \sigma_2 \laction x_1 = i \; x_2, \;\;\sigma_3 \laction x_1 = x_1, \nn \\
	& & \sigma_1 \laction x_2 = x_1, \;\; \sigma_2 \laction x_2 = - i \; x_1, \;\;\sigma_3 \laction x_2 = - x_2 \nn
\eeqarr
The Hopf-algebra $U(sl_2)$ thus gets represented in the accordingly dimensioned Hopf-algebra of vector fields 
$\mathfrak{W}(\Pi,\X)$ by
\beqarr
 \sigma_1 & = & i ( \mathfrak{w}^0_{2 1} + \mathfrak{w}^0_{1 2}) \nn \\
 \sigma_2 & = & \mathfrak{w}^0_{1 2} - \mathfrak{w}^0_{2 1} \nn \\
 \sigma_3 & = & i ( \mathfrak{w}^0_{1 1} - \mathfrak{w}^0_{2 2} ) \nn
\eeqarr
For the twist-deformation of these coproducts we now merely have to insert these expressions in those for the
coproducts of $\sigma_i$ from above and afterwards insert the twisted expressions for the vector fields from 
the last section. In particular for the twist (\ref{htwist}) we obtain in two dimensions the following explicit expressions for the twisted coproducts of $\mathfrak{w}^+_1$ and $\mathfrak{w}^+_2$ to be 
\beqarr
	\coproduct_{\twist_h}(\mathfrak{w}^+_1) & = & 
		\mathfrak{w}^+_1 \tensor \one + \one \tensor \mathfrak{w}^+_1
		+ \mathfrak{w}^+_1 \tensor (e^{- h \mathfrak{w}^+_2} - 1)\nn \\
	\coproduct_{\twist_h}(\mathfrak{w}^+_2) & = & 
		\mathfrak{w}^+_2 \tensor \one + \one \tensor \mathfrak{w}^+_2 .\nn
\eeqarr
We as well obtain the twisted coproducts of $\mathfrak{w}^0_{1 1}$, $\mathfrak{w}^0_{1 2}$, 
$\mathfrak{w}^0_{2 1}$ and $\mathfrak{w}^0_{2 1}$ to be given by
\beqarr
	\coproduct_{\twist_h}(\mathfrak{w}^0_{1 1}) & = & 
		\mathfrak{w}^0_{1 1} \tensor \one + \one \tensor \mathfrak{w}^0_{1 1}\nn \\
	\coproduct_{\twist_h}(\mathfrak{w}^0_{1 2}) & = &
		\mathfrak{w}^0_{1 2} \tensor \one + \one \tensor \mathfrak{w}^0_{1 2}
		+ \mathfrak{w}^0_{1 2} \tensor (e^{+ h \mathfrak{w}^+_2} - 1) \nn \\
	\coproduct_{\twist_h}(\mathfrak{w}^0_{2 1}) & = & 
		\mathfrak{w}^0_{2 1} \tensor \one + \one \tensor \mathfrak{w}^0_{2 1}
		+ h \; \mathfrak{w}^0_{1 1} \tensor \mathfrak{w}^+_1
		+ \mathfrak{w}^0_{2 1} \tensor (e^{- h \mathfrak{w}^+_2} - 1) \nn \\
	\coproduct_{\twist_h}(\mathfrak{w}^0_{2 2}) & = & 
		\mathfrak{w}^0_{2 2} \tensor \one + \one \tensor \mathfrak{w}^0_{2 2}
		+ h \; \mathfrak{w}^0_{1 1} \tensor \mathfrak{w}^+_2 .\nn
\eeqarr
The generating relations of $U(\X_{\twist_h})$ then read
$$
	x_1 *_{\twist_h} x_2 - x_2 *_{\twist_h} x_1 = i \; h \; x_1.
$$
The twisted coproducts of the generators $\sigma_i$ of $U_{\twist_h}(sl_2)$ are then given by
\beqarr
 \coproduct_{\twist_h}(\sigma_1) 
 	& = & i ( \coproduct_{\twist_h}(\mathfrak{w}^0_{2 1}) 
 	+ \coproduct_{\twist_h}(\mathfrak{w}^0_{1 2})) \nn \\
 \coproduct_{\twist_h}(\sigma_2) 
 	& = & \coproduct_{\twist_h}(\mathfrak{w}^0_{1 2}) 
 	- \coproduct_{\twist_h}(\mathfrak{w}^0_{2 1}) \nn \\
 \coproduct_{\twist_h}(\sigma_3) 
 	& = & i ( \coproduct_{\twist_h}(\mathfrak{w}^0_{1 1}) 
 	- \coproduct_{\twist_h}(\mathfrak{w}^0_{2 2}) ). \nn
\eeqarr

\section{Closing Remarks}

We introduced a general construction that allows for an introduction of a Hopf-algebra
of vector fields on a finitely generated representation space of universal enveloping algebra type.
Existing representations of $U(\liealgebra)$ can be embedded into the vector fields. Since the latter 
is larger than $U(\liealgebra)$, twisting of the vector fields provides a larger varity of deformations for
$U(\liealgebra)$ that could not be obtained within $U(\liealgebra)$ alone. In the mean time the twists
of our vector fields are nothing else than starproducts. In the last section we presented some examples 
that outline applicability of our construction. However, we emphasize that this setup is of course not 
restricted to commutative vector fields as the examples might suggest.

\section{Acknowledgment}

A this point we would like to thank Prof. Dr. Julius Wess for his wise support, trust and patience.
We moreover thank Dr. Branislav Jur\v{c}o, Dr. Alexander Schmidt and Dr. Hartmut Wachter for 
fruitfull discussions and carefully reading the manuscript.


\begin{thebibliography}{10}

\bibitem{Aschieri:2005zs}
P.~Aschieri, M.~Dimitrijevic, F.~Meyer, and J.~Wess.
\newblock Noncommutative geometry and gravity.
\newblock {\em Class. Quant. Grav.}, 23:1883--1912, 2006.

\bibitem{Aschieri:2005yw}
P.~Aschieri et~al.
\newblock A gravity theory on noncommutative spaces.
\newblock {\em Class. Quant. Grav.}, 22:3511--3532, 2005.

\bibitem{Blohmann:2002wx}
C.~Blohmann.
\newblock Covariant realization of quantum spaces as star products by drinfeld
  twists.
\newblock {\em J. Math. Phys.}, 44:4736--4755, 2003.

\bibitem{Borowiec:2005bq}
A.~Borowiec, J.~Lukierski, and V.~N. Tolstoy.
\newblock On twist quantizations of d = 4 lorentz and poincare algebras.
\newblock {\em Czech. J. Phys.}, 55:11, 2005.

\bibitem{Chaichian:2004za}
M.~Chaichian, P.~P. Kulish, K.~Nishijima, and A.~Tureanu.
\newblock On a lorentz-invariant interpretation of noncommutative space-time
  and its implications on noncommutative qft.
\newblock {\em Phys. Lett.}, B604:98--102, 2004.

\bibitem{Charipressley}
V.~Chari and A.~N. Pressley.
\newblock {\em A Guide to Quantum Groups}.
\newblock Cambridge University Press, 1998.

\bibitem{Chu:1998qz}
C.-S. Chu and P.-M. Ho.
\newblock Noncommutative open string and d-brane.
\newblock {\em Nucl. Phys.}, B550:151--168, 1999.

\bibitem{Jambor:2004kc}
C.~Jambor and A.~Sykora.
\newblock Realization of algebras with the help of *-products.
\newblock {\em hep-th/0405268}, 2004.

\bibitem{Kempf:1994qp}
A.~Kempf.
\newblock Quantum field theory with nonzero minimal uncertainties in positions
  and momenta.
\newblock {\em hep-th/9405067}, 1994.

\bibitem{Kempf:1997sr}
A.~Kempf.
\newblock Recent results on uv-regularisation through uv-modified uncertainty
  relations.
\newblock {\em hep-th/9711204}, 1997.

\bibitem{Kempf:1998gk}
A.~Kempf.
\newblock On the structure of space-time at the planck scale.
\newblock {\em hep-th/9810215}, 1998.

\bibitem{Kempf:1998em}
A.~Kempf.
\newblock On the three short-distance structures which can be described by
  linear operators.
\newblock {\em Rept. Math. Phys.}, 43:171--177, 1999.

\bibitem{Koch:2004ud}
F.~Koch and E.~Tsouchnika.
\newblock Construction of theta-poincare algebras and their invariants on
  m(theta).
\newblock {\em Nucl. Phys.}, B717:387--403, 2005.

\bibitem{Lukierski:1991ff}
J.~Lukierski, A.~Nowicki, and H.~Ruegg.
\newblock Real forms of complex quantum anti-de sitter algebra u- q(sp(4:c))
  and their contraction schemes.
\newblock {\em Phys. Lett.}, B271:321--328, 1991.

\bibitem{Lukierski:1992dt}
J.~Lukierski, A.~Nowicki, and H.~Ruegg.
\newblock New quantum poincare algebra and k deformed field theory.
\newblock {\em Phys. Lett.}, B293:344--352, 1992.

\bibitem{Lukierski:1991pn}
J.~Lukierski, H.~Ruegg, A.~Nowicki, and V.~N. Tolstoi.
\newblock Q deformation of poincare algebra.
\newblock {\em Phys. Lett.}, B264:331--338, 1991.

\bibitem{Lukierski:2005jb}
J.~Lukierski and M.~Woronowicz.
\newblock Twisted space-time sym\-metry, non-com\-mu\-ta\-tivity and particle dynamics.
\newblock {\em hep-th/0512046}, 2005.

\bibitem{Lukierski:2005fc}
J.~Lukierski and M.~Woronowicz.
\newblock New lie-algebraic and quadratic deformations of minkowski space from
  twisted poincare symmetries.
\newblock {\em Phys. Lett.}, B633:116--124, 2006.

\bibitem{Majid}
S.~Majid.
\newblock {\em Foundations of Quantum Group Theory}.
\newblock Cambridge University Press, 2000.

\bibitem{Majid:1998up}
S.~Majid and R.~Oeckl.
\newblock Twisting of quantum differentials and the planck scale hopf algebra.
\newblock {\em Commun. Math. Phys.}, 205:617--655, 1999.

\bibitem{Majid:1994cy}
S.~Majid and H.~Ruegg.
\newblock Bicrossproduct structure of kappa poincare group and noncommutative
  geometry.
\newblock {\em Phys. Lett.}, B334:348--354, 1994.

\bibitem{Oeckl:2000eg}
R.~Oeckl.
\newblock Untwisting noncommutative r**d and the equivalence of quantum field
  theories.
\newblock {\em Nucl. Phys.}, B581:559--574, 2000.

\bibitem{Schomerus:1999ug}
V.~Schomerus.
\newblock D-branes and deformation quantization.
\newblock {\em JHEP}, 06:030, 1999.

\bibitem{Seiberg:1999vs}
N.~Seiberg and E.~Witten.
\newblock String theory and noncommutative geometry.
\newblock {\em JHEP}, 09:032, 1999.

\bibitem{Wess:2003da}
J.~Wess.
\newblock Deformed coordinate spaces: Derivatives.
\newblock {\em hep-th/0408080}, 2003.

\bibitem{Zachos:2001ux}
C.~K. Zachos.
\newblock Deformation quantization: Quantum mechanics lives and works in
  phase-space.
\newblock {\em Int. J. Mod. Phys.}, A17:297--316, 2002.

\end{thebibliography}
\end{document}